\documentclass[a4paper,10pt]{article}

\usepackage{graphicx}
\usepackage{amsmath}
\usepackage{citesort}
\usepackage[sort&compress,numbers]{natbib}
\usepackage{amsfonts}
\usepackage{amssymb}
\usepackage{amsmath}
\usepackage{latexsym}
\usepackage{color}
\usepackage{ntheorem}
\usepackage{braket}
\usepackage{url}

\newcommand{\Bo}{{\mbox{\rm\scriptsize Bo}}}
\newcommand{\Bobo}{{\mbox{\rm\tiny Bo}}}

\newcommand{\mTB}{m_{\mbox{\rm\scriptsize TB}} }

\newcounter{mnotecount}[section]

\renewcommand{\themnotecount}{\thesection.\arabic{mnotecount}}
\newcommand{\mnote}[1]
{\protect{\stepcounter{mnotecount}}$^{\mbox{\footnotesize
$
\bullet$\themnotecount}}$ \marginpar{
\raggedright\tiny\em
$\!\!\!\!\!\!\,\bullet$\themnotecount: #1} }

\newcommand{\ol}[1]{\overline{#1}{}}

\newcommand{\jlcax}[1]{}
\newcommand{\eean}{\nonumber\end{eqnarray}}

\newcommand{\og}{{\overline{g}}}

\newcommand{\kk}[1]{}

\newcommand{\beq}{\begin{equation}}

\newcommand{\FS}       
                  {F}

\newcommand{\HS} 
       {H_{\mbox{\scriptsize volume}}}

\newcommand{\half}{\frac 12}

{\ptc{this should be removed in the oberwolfach version}}%

\newcommand{\eeal}[1]{\label{#1}\end{eqnarray}}
\newcommand{\bed}{\begin{deqarr}}
\newcommand{\eed}{\end{deqarr}}
\newcommand{\bedl}[1]{\begin{deqarr}\label{#1}}
\newcommand{\eedl}[2]{\arrlabel{#1}\label{#2}\end{deqarr}}

\newcommand{\mcN}{{\mycal N}}

\newcommand{\bel}[1]{\begin{equation}\label{#1}}
\newcommand{\bea}{\begin{eqnarray}}
\newcommand{\bean}{\begin{eqnarray}\nonumber}
\newcommand{\beal}[1]{\begin{eqnarray}\label{#1}}
\newcommand{\eea}{\end{eqnarray}}

\newcommand{\Eq}[1]{Equation~\eq{#1}}

\newcommand{\Eqs}[2]{Equations~\eq{#1}-\eq{#2}}

\newcommand{\be}{\begin{equation}}
\newcommand{\eeq}{\end{equation}}
\newcommand{\ee}{\end{equation}}
\newcommand{\beqa}{\begin{eqnarray}}
\newcommand{\eeqa}{\end{eqnarray}}
\newcommand{\beqan}{\begin{eqnarray*}}
\newcommand{\eeqan}{\end{eqnarray*}}
\newcommand{\ba}{\begin{array}}
\newcommand{\ea}{\end{array}}

\newcommand{\scri}{{\mycal I}}%
\newcommand{\scrip}{\scri^{+}}%

\newcommand{\warn}[1]
{\protect{\stepcounter{mnotecount}}$^{\mbox{\footnotesize
$
\bullet$\themnotecount}}$ \marginpar{
\raggedright\tiny\em
$\!\!\!\!\!\!\,\bullet$\themnotecount: {\bf Warning:} #1} }

\newcommand{\R}{\mathbb R}

\newcommand{\eq}[1]{(\ref{#1})}

\newcommand{\ptc}[1]{\mnote{{\bf ptc:}#1}}

\newcommand{\beqar}{\begin{deqarr}}
\newcommand{\eeqar}{\end{deqarr}}

\newcommand{\beaa}{\begin{eqnarray*}}
\newcommand{\eeaa}{\end{eqnarray*}}

\DeclareFontFamily{OT1}{rsfs}{}
\DeclareFontShape{OT1}{rsfs}{m}{n}{ <-7> rsfs5 <7-10> rsfs7 <10-> rsfs10}{}
\DeclareMathAlphabet{\mycal}{OT1}{rsfs}{m}{n}

\begin{document}

\title{The mass of light-cones%
\thanks{Preprint UWThPh-2014-2.  Supported in part by the Austrian Science Fund (FWF): P 24170-N16. PTC is grateful to the Stanford Mathematics Research Center for hospitality and support during work on this paper.}}
\author{
Piotr T. Chru\'sciel{}\thanks{URL \protect\url{homepage.univie.ac.at/piotr.chrusciel}, Email  {piotr.chrusciel@univie.ac.at}}\ \ and
Tim-Torben Paetz{}\thanks{Email  Tim-Torben.Paetz@univie.ac.at}   
 \vspace{0.5em}\\  \textit{Gravitational Physics, University of Vienna}  \\ \textit{Boltzmanngasse 5, 1090 Vienna, Austria }}
\maketitle

\vspace{-0.2em}

\begin{abstract}
We give an elementary proof of positivity of total gravitational energy in space-times containing complete smooth light-cones.
\end{abstract}

\noindent
\hspace{2.1em} PACS: 04.20.Cv, 04.20.Ex, 04.20.Ha

\bigskip

One of the deepest questions arising in general relativity is that of positivity of total energy. Years of attempts by many authors have led to an affirmative answer in the milestone papers of Schoen and Yau~\cite{schoen:yau:ADM,SchoenYau81} and Witten~\cite{Witten81}; see also~\cite{CJL,LudvigsenVickers82,schoen:yau:bondi}. These proofs use sophisticated PDE techniques, with positivity resulting from the analysis of solutions of seemingly unrelated partial differential equations. The aim of this letter is to show that an elementary direct proof of positivity can be given for a large class of space-times, namely those containing globally smooth light-cones. As a bonus, our proof gives an explicit positive-definite expression for the mass,
equation \eq{29XII13.16+},
 in terms of physically relevant fields, such as the shear of the light-cone.

Thus,  consider a globally smooth, null-geodesically complete light-cone in an asymptotically Minkowskian space-time.
The formula \eq{Bondi_mass_dfn} below for the mass in this context has been derived by  Bondi et al.~\cite{BBM,Sachs}, compare~\cite{T}. We show how to rewrite this formula in terms of geometric data on the light-cone, Equation~\eq{29XII13.11} below.
The constraint equations induced by Einstein's field equations on the light-cone
 are then used to obtain our manifestly positive mass formula \eq{29XII13.16+} by elementary manipulations.

The initial  data on the light-cone comprise a pair $(\mcN,\check g)$,
where $\mcN=\R^3\setminus \{0\}$ and  $\check g$ is a smooth field of symmetric two-covariant tensors on $\mcN$ of signature $(0,+,+)$ such that $\check g(\partial_r,\cdot)=0$.%
\footnote{To avoid an ambiguity in notation we write  $\check g$ for what was denoted by ${\tilde g}$ in~\cite{CCM2}, as ${\tilde g}$ is usually used for the  conformally rescaled metric when discussing $\scrip$.}
 The vertex $O$ of the light-cone $C_O:=\mcN \cup \{O\}$
is located at the origin of $\R^3$, and the half-rays issued from the origin correspond to the generators of $C_O$. For simplicity we assume throughout that the initial data lead to a smooth space-time metric, cf.~\cite{ChConeExistence}. The requirements of regularity at the origin,  asymptotic flatness, and global smoothness, lead to the following restrictions on $\check g$:

Letting $(r,x^A)$, $A\in\{2,3\}$, denote spherical coordinates on $\R^3$, and writing $s_{AB}dx^A dx^B$ for the unit round metric on $S^2$, regularity conditions at the vertex imply that the coordinate $r$ can be chosen so that for small $r$  we have
    \beal{29XII13.1}
    &
     \check g \equiv \og_{AB}dx^A dx^ B = r^2( s_{AB} + h_{AB})dx^A dx^ B
     \;,
     &
     \\
    \label{29XII13.2-}
    &
      h_{AB}   = O(r^2)
     \;,\
     \partial_C h_{AB}   = O(r^2)
     \;,
     \
     \partial_r h_{AB}   = O(r )
    \;,
     &
    \eeal{29XII13.2}
see~\cite[Section~4.5]{CCM2} for a detailed discussion, including properties of higher derivatives; the latter are assumed implicitly whenever needed below.
Here, and elsewhere, an overline denotes a space-time object restricted to the light-cone.

   Existence of a Penrose-type conformal completion
implies that
the coordinate $r$ can be chosen so that for large $r$ we have
    \beal{29XII13.3}
    &
     \og_{AB}  = r^2 (\og_{AB})_{-2} +r (\og_{AB})_{-1} + \psi_{AB}
     \;,
     &
     \\
    \label{29XII13.4-}
    &
      \psi_{AB}   = O(1)
     \;,\
     \partial_C \psi_{AB}   = O(1)
     \;,
     \
     \partial_r \psi_{AB}   = O(r^{-2})
     \;,
     &
     \\
    &
       \partial_r^2 \psi_{AB}   = O(r^{-3})
\;,
\
     \partial_C \partial_D \psi_{AB}   = O(1)
     \;,
     &
     \\
    &
          \partial_r \partial_C \psi_{AB}   = O(r^{-2} )
     \;,
      \
          \partial_r \partial_C \partial_D\psi_{AB}   = O(r^{-2} )
     \;,
     &
\\
&
          \partial_r^2 \partial_C \psi_{AB}   = O(r^{-3} )
     \;,
      \
          \partial_r^2 \partial_C \partial_D\psi_{AB}   = O(r^{-3} )
     \;,
     &
    \eeal{29XII13.4}
    for some smooth tensors $(\og_{AB})_{-i}=(\og_{AB})_{-i}(x^C)$, $i=1,2$, on $S^2$.
%
%

\Eqs{29XII13.3}{29XII13.4} are necessary for existence of a smooth $\scrip$, but certainly not sufficient: our conditions admit initial data sets which might lead to a polyhomogeneous but \emph{not} smooth $\scrip$; see~\cite{ChPaetz3,TimAsymptotics,andersson:chrusciel:PRL,ChMS}.

We denote by $\tau$ the \emph{divergence}, sometimes called \emph{expansion}, of the light-cone:
    \bel{29XII13.5}
     \tau := \chi_{A}{}^{A}
     \;,
     \
     \mbox{where}
     \
      \chi_{A}{}^{B}:=\half    \og^{BC} \partial_r \og_{AC}
     \;.
    \ee
Conditions \eq{29XII13.1}-\eq{29XII13.4} imply
\begin{eqnarray}
\tau &=& 2r^{-1} + \tau_2 r^{-2} + O(r^{-3}) \text{ for large $r$}
\label{beh_tau_large}
\;,
\\
\tau &=& 2r^{-1} + O(r)  \text{ for small $r$}
\label{beh_tau_small}
\;.
\end{eqnarray}
In particular $\tau $ is positive in both regions.
    Now, standard arguments show that if  $\tau$  becomes negative somewhere, then the light-cone will either fail to be globally smooth, or the space-time will not be null-geodesically complete. So our requirement of global smoothness of the light-cone together  with completeness of generators imposes the condition%
    \footnote{The Raychaudhuri equation \eq{10XII13.2} below with $\kappa=0$ implies that $\tau$ is monotonous non-increasing, which yields \eq{29XII13.6} directly in any case after noting  that the sign of $\tau$ is invariant under orientation-preserving changes of parametrisation of the generators.}
    \bel{29XII13.6}
     \tau   >0
     \;.
    \ee
We further require
\begin{equation}
 \det (\og_{AB})_{-2}>0
  \;,
\label{Riemannian}
\end{equation}
which excludes conjugate points at the intersection of the light-cone with $\scrip$.
    Both conditions will be assumed to hold from now on.

    The inequality \eq{29XII13.6} implies that the connection coefficient $\kappa$,
     defined through the equation
     $$
     \nabla_{\partial_r}\partial_r = \kappa \partial_r
     \;,
     $$
and measuring thus how the parameter $r$ differs from an affine-one, can be algebraically calculated from the Raychaudhuri equation:
  \bean
   \lefteqn{
    \partial_r\tau - \kappa \tau+ \frac{\tau^2}{2} + |\sigma|^2 + 8 \pi T_{rr}|_\mcN=0
    }
     &&
     \\
     &&
    \Longleftrightarrow
    \quad
     \kappa = \frac 1 \tau \big( \partial_r\tau + \frac{\tau^2}{2} + |\sigma|^2 + 8 \pi T_{rr}|_\mcN
    \big)
  \;.
  \eeal{10XII13.2}
  Here   $\sigma$ is the shear of the light-cone:
  \bel{29XII13.6w}
   \sigma_{A}{}^{B}= \chi_{A}{}^{B} - \half  \tau \delta_A{}^B
   \;,
   \ee
which satisfies
\begin{equation}
 \sigma_A{}^B =O(r) \text{ for small $r$}\;, \enspace   \sigma_A{}^B =O(r^{-2}) \text{ for large $r$}\;.
\label{beh_sigma}
\end{equation}
   Assuming that
   \bel{29XII13.8}
    T_{rr}|_\mcN=O(r^{-4}) \text{ for large $r$}
    \;,
    \ee
   we find from \eq{10XII13.2} and our previous hypotheses
   \bel{29XII13.7}
    \mbox{$\kappa = O(r)$ for small $r$\;, \enspace $\kappa = O(r^{-3})$ for large $r$.}
   \ee

For the proof of positivity of the Trautman-Bondi mass it will be convenient to change $r$ to a new coordinate so that $(\overline g_{AB})_{-2}$ in \eq{29XII13.3} is the unit round sphere metric and that the resulting $\kappa$ vanishes (i.e.\ the new coordinate $r$ will be an affine parameter along the generators of the light-cone).
\newcommand{\nc}{r_{\mathrm{as}}}%
Denoting momentarily the new coordinate by $\nc$, and using the fact that every metric on $S^2$ is conformal to the unit round metric $s_{AB}$, the result  is achieved by setting
\begin{eqnarray}
  \nc(r,x^A) &=& \Theta(x^A) \int_0^r e^{H(\hat r,x^A)} d\hat r
 \;,
\label{dfn_mathring_varphi}
\\
 H(r,x^A) &=&- \int_r ^\infty  \kappa(\tilde r,x^A)   d\tilde r \;,
\\
\Theta(x^A) &=& \Big(\frac{\det (\overline g_{AB})_{-2}}{\det s_{AB}}\Big)^{1/4}
\;.
\end{eqnarray}
The functions $r\mapsto \nc (r,x^A) $ are strictly increasing with $\nc (0,x^A)=0$. \Eq{29XII13.7} shows that there exists a constant $C$ such that for all $r$ we have
\bel{9XII13.1}
e^{-C}\le e^H \le e^ C
  \;,
\ee
which  implies that $\lim_{r\rightarrow \infty} \nc (r,x^A)=+\infty$. We conclude that for each $x^A$ the function $r\mapsto \nc (r,x^A)$ defines a smooth
bijection from $\mathbb{R}^+$ to itself.
Consequently,  smooth inverse functions $\nc \mapsto r(\nc ,x^A)$ exist.

We have normalised the affine parameter $\nc$ so that 
\begin{eqnarray}
 \nc (r,x^A)&=&\Theta(x^A) r +(\nc)_\infty (x^A) + O(r^{-1})
 \;,
\label{29XII13.9}
\end{eqnarray}
for large $r$, which implies
\begin{eqnarray}
 \nc (r,x^A) &=&   (\nc)_0(x^A)r + O(r^3) \text{ for small $r$}
 \;,
\label{trafo_small}
\end{eqnarray}
where
\begin{eqnarray}
  (\nc)_\infty (x^A)
   & = &
\Theta(x^A) \int_0^\infty\big( e^{H( r, x^A)}- 1\big)dr
\;,
\label{29XII13.10}
\\
(\nc)_0(x^A) &=&   \Theta(x^A) e^{- \int_0 ^\infty  \kappa(\tilde r,x^A)   d\tilde r}
 \;.
\end{eqnarray}
After some obvious redefinitions, for  $\nc$  large  \eq{29XII13.3} becomes
\begin{equation}
      \og_{AB}  = \nc^2 s_{AB} +\nc (\og_{AB})_{-1} + \psi_{AB}\;.
\label{standard_form}
\end{equation}
The boundary conditions \eq{29XII13.4-}-\eq{29XII13.4} remain unchanged when $r$ is replaced by $\nc$ there.
 On the other hand,  \eq{29XII13.1}-\eq{29XII13.2} will not be true anymore.
However, we note for further use that the coordinate transformation \eq{dfn_mathring_varphi} preserves the behavior of $\tau$ and $\sigma$ near the vertex. Indeed, inserting $r=r(\nc)$ into \eq{29XII13.3} and using the definitions \eq{29XII13.5} and \eq{29XII13.6w} one finds that \eq{beh_tau_large}-\eq{beh_tau_small} and \eq{beh_sigma} continue to hold with $r$ replaced by $\nc$.

A key role in what follows will be played by   the equation~\cite[Equations (10.33)  and (10.36)]{CCM2},%
\footnote{On the right-hand-side of the second equality  in \cite[Equation~(10.36)]{CCM2} a term $\tau \overline g^{11}/2$ is missing.}%
\begin{equation}
 (\partial_r + \tau + \kappa)\zeta +  \check R - \frac{1}{2}|\xi|^2 + \overline g^{AB}\tilde\nabla_A\xi_B = S
 \;,
 \label{zeta_constraint}
\end{equation}
where $
 |\xi|^2 := \overline g^{AB}\xi_A\xi_B$.
In coordinates adapted to the light-cone as in \cite{CCM2} the space-time formula for the auxiliary function $\zeta$ is
\bel{30XII13.2}
 \zeta=\big(2   g^{AB} \Gamma^r_{AB} + \tau  g^{rr}\big)\big|_{\mcN}
  \;,
\ee
and $\zeta$ is in fact the divergence of the family  of suitably normalized null generators normal to the spheres of constant $r$ and  transverse to $\mcN$. Here $\tilde \nabla$ denotes the Levi-Civita connection of $\check g$ 
viewed as a metric on $S^2$ (more precisely, an $r$-dependent family of metrics).
The symbol $\check R$ denotes the curvature scalar of $\check g$.
The connection coefficients
$\xi_A\equiv -2  \Gamma^r_{rA}|_{\mcN}$ are determined by \cite[Equation (9.2)]{CCM2}:
\be
  \frac{1}{2}(\partial_r + \tau)\xi_A  - \tilde\nabla_B \sigma_A^{\phantom{A}B} + \frac{1}{2}\partial_A\tau +\partial_A \kappa
= - 8 \pi T_{rA}|_{\mcN}
 \;.
 \label{eqn_nuA_general}
\ee
Finally,
\begin{eqnarray}
 \nonumber
 S &:= & 8 \pi ( g^{AB}T_{AB}- g^{\mu\nu}T_{\mu\nu})|_{\mcN}
\\
 &= &-8 \pi \big(
  {g}^{rr} {T}_{rr}+2 {g}^{rA} {T}_{rA}
 +2 {g}^{ur} {T}_{ur}
  \big)|_\mcN
  \;,
\label{C0_3}
\end{eqnarray}
with $T_{ur}=T(\partial_u,\partial_r)$, where $\partial_u$ is transverse to $\mcN$. The first equality makes it clear that $S$ does not depend upon the choice of coordinates away from $\mcN$. In a coordinate system where $\overline g^{rr}=\overline g^{rA}=0$ we have  $S= -16 \pi g^{ur} T_{ur}|_\mcN$ which, with our signature $(-,+,+,+)$, is non-negative for matter fields satisfying the dominant energy condition when both $\partial_r$ and $\partial_u$ are causal future pointing, as will be assumed from now on.
%
%

Letting $d\mu_{{\check g}}=\sqrt{\det \og_{AB}}dx^2dx^3$,
we derive below the following surprising formula for the Trautman-Bondi~\cite{BBM,Sachs,T} mass $\mTB$ of complete light-cones:
\begin{equation}
 \mTB
  = \frac 1 {  16 \pi}
 \int_0^\infty  \int_{S^2} \bigg(  \half   |\xi|^2  + S
   +
(  |\sigma|^2
    +
  8 \pi T_{rr}|_{\mcN}
   )e^{   \int_r^\infty \frac {\tilde r\tau-2} {2 \tilde r} \,d\tilde r  } \bigg)
   d\mu_{{\check g}} dr
   \!\!
 \;.
\label{29XII13.16+}
\end{equation}
The coordinate $r$ here is an affine parameter along the generators
normalised so that $r=0$ at the vertex, with \eq{standard_form} holding for large $r$.
Positivity of $\mTB $ obviously follows in vacuum. For matter fields satisfying the dominant energy condition we have $S\ge 0$, $T_{rr}\ge 0$  and positivity again follows.

Note that since $\mTB$  decreases when sections of $\scrip$ are moved to the future,
\eq{29XII13.16+} provides an a priori bound on the integrals appearing there both on $\mcN$ and for all later light-cones, which is likely to be useful when analysing the global behaviour of solutions of the Einstein equations.

To prove \eq{29XII13.16+},  we assume \eq{29XII13.8}.
We change coordinates via \eq{dfn_mathring_varphi}, and use now the symbol $r$ for the coordinate $\nc$. Thus $\kappa=0$, we have \eq{29XII13.3}  with $(g_{AB})_{-2}=s_{AB}$, and further \eq{29XII13.4-}-\eq{29XII13.4},
\eq{beh_tau_large}-\eq{beh_tau_small} and \eq{beh_sigma} hold. For $r$ large  one immediately obtains
\beal{29XII13.12}
 \sqrt{\det \og_{AB}} &= &  r^{2} \sqrt{\det s_{AB}}
  \big( 1- \tau_2   r^{-1} + O(r^{-2})\big)
 \;.
\eeal{29XII13.13}
Let us further assume that, again for large $r$,
\begin{equation}
     T_{rA}|_\mcN=O(r^{-3}) \;, \quad    S = O(r^{-4})
\;.
\label{29XII13.17}
\end{equation}
It then follows from \eq{eqn_nuA_general} and our remaining hypotheses that  $\xi_A$
 satisfies
\bel{29XII13.3x}
 \xi_A = (\xi_A)_1 r^{-1} + o(r^{-1}) \text{ and } \partial_B\xi_A = O(r^{-1})\;,  
\end{equation}
for some smooth covector field $(\xi_A)_1$ on $S^2$. An analysis of \eq{zeta_constraint} gives
\bea
 \zeta(r,x^A)
  & = &
    -2r^{-1} + \zeta_2 (x^A) r^{-2} + o(r^{-2})
 \;,
\eeal{29XII13.13+}
with a smooth function $\zeta_2$.
Regularity at the vertex requires that for small $r$
\bel{29XII13.21}
\xi_A=O(r^2)\;,
\quad
 \zeta =  O(r^{-1})
 \;,
\ee
where $r$ is the original coordinate which makes the initial data manifestly regular at the vertex.
Using the transformation formulae for connection coefficients one checks  that this behavior is preserved under \eq{trafo_small}.

We will show shortly
that \emph{if}
the light-cone data arise from a space-time with a smooth conformal completion at null infinity $\scrip$, and \emph{if} the light-cone intersects $\scrip$ in a smooth cross-section $S$, then  the Trautman-Bondi mass of $S$ equals
\bel{29XII13.11}
 \mTB    =\frac{1}{16\pi}\int_{S^2}( \zeta_2+ \tau_2) d\mu_s
 \;,
\ee
where $d\mu_s = \sqrt{\det s_{AB}} \,dx^2dx^3$.
Note that this justifies the use of \eq{29XII13.11} as the \emph{definition} of mass of an initial data set on a light-cone with complete generators, regardless of any space-time assumptions.

It follows from \eq{beh_tau_large}, \eq{29XII13.12} and \eq{29XII13.13+} that for large $r$ we have
\bean
 \int_{S^2} \zeta d\mu_{{\check g}}
 & = &
 \int_{S^2} ( - 2 r + \zeta_2 +o(1))(1 - \tau_2 r^{-1} +O(r^{-2}))  d\mu_{s}
\\
 & = &
  - 8 \pi r + \int_{S^2}   (\zeta_2  + 2\tau_2 )  d\mu_{s} + o(1)
 \;.
\eeal{28XII13.12}
This allows us to rewrite \eq{29XII13.11} as
\bel{29XII13.16}
 16 \pi \mTB
 = \lim _{r\to\infty}  \big( \int_{S^2} \zeta d\mu_{{\check g}} + 8 \pi r\big) - \int_{S^2} \tau_2 d\mu_s
 \;.
\ee
To establish \eq{29XII13.16+}, first
note that from (\ref{zeta_constraint}) with $\kappa=0$ and the Gauss-Bonnet theorem we have,
using  $\partial_r\sqrt{\det \overline g_{AB}} = \tau \sqrt{\det \overline g_{AB}}$,
\bel{28XII13.10}
 \partial_r \int_{S^2} \zeta d\mu_{{\check g}}
  =
  -8 \pi + \int_{S^2} \big(  \half  |\xi|^2  +S\big) d\mu_{{\check g}}
 \;.
\ee
Integrating in $r$ and using \eq{29XII13.17}-\eq{29XII13.21}  one obtains
\bean
     \lim_{r\to\infty} \big(
   \int_{S^2} \zeta   d\mu_{{\check g}} + 8 \pi r\big)
 & = &
    \int_0^\infty \int_{S^2}  \big(\half |\xi|^2 +S)  d\mu_{{\check g}} dr
 \;.
\eeal{28XII13.11}

Next,
let $\tau_1:=2/r$, $\delta \tau := \tau - \tau_1$.
It follows from the Raychaudhuri equation with $\kappa=0$ that $\delta \tau$ satisfies the equation
$$
 \frac{  {d}\delta \tau } {{d}r} + \frac {\tau+\tau_1} 2 \delta \tau = -|\sigma|^2 - 8 \pi T_{rr}|_\mcN
 \;.
$$
Letting
%
\bel{9XII13.9}
 \Psi =  \exp\big(  \int_0^r \frac {\tilde r\tau-2} {2 \tilde r} \,d\tilde r \big)
 \;,
\ee
and using \eq{beh_tau_large}-\eq{beh_tau_small} and \eq{beh_sigma}-\eq{29XII13.8}
one  finds
\bean
 \delta \tau (r ) & = & -r^{-2}\Psi^{-1}   \int_{0}^{r} (  |\sigma|^2 +  8 \pi T_{rr}|_\mcN )\Psi r^2 dr
\\
 & = &  {\tau_2 }{r^{-2}} +o(r^{-2})
 \;,
\eeal{5I14.1}
where
\bel{9XII13.10}
 \tau_2 = - \lim_{r\to\infty}  {\Psi^{-1}}  \int_0^r  (  |\sigma|^2 +
  8 \pi T_{rr}|_\mcN
   )\Psi r^2 dr
    \le 0
 \;.
\ee
Inserting this into \eq{29XII13.16} gives \eq{29XII13.16+} after noting that
\begin{equation}
d\mu_{\check g}=e^{  - \int_r^\infty \frac {\tilde r\tau-2} { \tilde r} \,d\tilde r  } r^2 d\mu_s
\;.
\label{volume_element}
\end{equation}

To continue, suppose that $\mTB  $ vanishes. It then follows from \eq{9XII13.10} that $T_{rr} |_\mcN=0=\sigma$. In vacuum this implies~\cite{CCG} that the metric is flat to the future of the light-cone. In fact, for many matter models the vanishing of $T_{rr}$ on the light-cone implies the vanishing of $T_{\mu\nu}$ to the future of the light-cone~\cite{CCG}, and the same conclusion can then be obtained.

It remains to establish \eq{29XII13.11}. We decorate with a symbol ``$\Bo$'' all fields arising in Bondi coordinates. Consider characteristic data in Bondi coordinates,  possibly defined only for large values of $r_\Bo$. The space-time metric on $\mcN=\{u ^\Bo=0\}$
can be written as
\bel{31XII13.1}
 \overline g = \overline g_{00}^\Bo {d}u_\Bo^2 + 2\nu_0^\Bo{d}u_\Bo {d}r_\Bo + 2\nu_A^\Bo{d}u _\Bo {d}x^A_\Bo + \check g^\Bo
 \;.
\ee
Under the usual asymptotic conditions on $\nu_A^\Bo$ and $\nu_0^\Bo$ one has (see, e.g., \cite{ChMS,Sachs})
\begin{equation}
 \og_{00}^\Bo =
  - 1 + \frac{2 M(x^A_\Bo)}{ r_\Bo}
+O(r_\Bo^{-2})
 \label{31XII13.7}
\;,
\end{equation}
and the Bondi mass is then defined as
\begin{eqnarray}
 \mTB &=& \frac{1}{4\pi}\int_{S^2} M \,d\mu_s
 \;.
\label{Bondi_mass_dfn}
\end{eqnarray}
In Bondi coordinates \eq{30XII13.2} becomes
\bel{30XII13.2+}
 \zeta^\Bo=
  2\Big( \frac{\check\nabla^A\nu_A^\Bo}{\nu^\Bo_0} -  \frac {(\og^\Bo)^{rr} }{r_\Bo}
  \Big)
  \;,
\ee
which allows us to express $(\og^\Bo)^{rr}$ in terms of $\zeta^\Bo$, leading eventually to
%
\bean
 \og_{00}^\Bo  &= &
  (\og^\Bo)^{AB}\nu_A^\Bo\nu_B^\Bo - (\nu_0^\Bo)^2(\og^\Bo)^{rr}
\\
 & = &
  - 1 + \frac{\zeta^\Bo_2 - 2 \mathring\nabla^A ( \nu_A^\Bo)_0}{2r_{\Bo}}
+O(r_\Bo^{-2})
\;,
\eeal{31XII13.5}
where $\mathring \nabla$ is the Levi-Civita connection of the metric $s_{AB}dx^Adx^B$, and $( \nu_A^\Bo)_0$ is the $r $-independent coefficient in an asymptotic expansion of $\nu_A^\Bo$.
Comparing with \eq{31XII13.7}, we conclude that
\bel{29XII13.22}
 \mTB    =\frac{1}{16\pi}\int_{S^2}  \zeta^\Bo_2 d\mu_s
 \;.
\ee
To finish the calculation we need to relate $\zeta^\Bo_2$ to the characteristic data.
In Bondi coordinates we have
$$
 \tau_\Bo = \frac 2 {r_\Bo}
  \;,
$$
and the Raychaudhuri equation implies that
$$
 \kappa^\Bo = r_\Bo \frac{|\sigma^\Bo|^2 + 8 \pi \overline T_{rr}} 2
  \;.
$$
The equation $\kappa\equiv\overline \Gamma {}^r_{rr}=0$ together with the usual transformation law for connection coefficients gives the equation
 \begin{eqnarray*}
 \lefteqn{
 \partial_{r_\Bo}\Big( \frac {\partial r_\Bo}{\partial r}\Big) + \kappa^\Bo \frac {\partial r_\Bo}{\partial r} = 0
  }
  &&
 \\
  &&
  \Longrightarrow
  \quad
  \frac{\partial r}{\partial r_\Bo} = e^{-\int_{r_\Bobo}^\infty \kappa^\Bobo}
   = 1 + O(r^{-2}_\Bo)
\;,
 \end{eqnarray*}
where we have used the asymptotic condition $\lim_{r\to\infty} \frac {\partial r_\Bo}{\partial r} = 1$. Hence
\begin{eqnarray*}
 r
  & = &
    \int_0^{r_\Bobo}   e^{-\int_{r_\Bobo}^\infty \kappa^\Bobo}
\\
     & = &
       r_\Bo + \underbrace{\int_0^{\infty}   \big(e^{-\int_{r_\Bobo}^\infty \kappa^\Bobo} -1\big)}_{=:-(r_\Bo)_{0}} + O(r^{-1}_\Bo)
 \;.
\end{eqnarray*}
To continue, we note that $x^A_\Bo=x^A$ and
$$
 \og^\Bo_{AB}(r_\Bo,x^A)=\og_{AB}\big(r(r_\Bo,x^A),x^A\big)
 \;,
$$
which implies
%
\bean
 \frac 2 {r_\Bo}
  & = &
  \tau^\Bo
= \tau   \partial_{r_\Bobo}r
  = \big(\frac 2 r + \frac{\tau_2}{r^2} +O(r^{-3})\big)\big(1 + O(r^{-2})\big)
\\
 \nonumber
 & = &
 \frac 2 r + \frac{\tau_2}{r^2} +O(r^{-3})
 \;,
\eeal{31XII13.8}
equivalently
\bel{31XII13.9}
 r_\Bo = r - \frac {\tau_2}2 + O(r^{-1})
 \
 \Longrightarrow
 \
 (r_\Bo)_0 =  -\frac {\tau_2}2
 \;.
\ee

We are ready now to transform  $\zeta  $ as given by \eq{30XII13.2}
to the new coordinate system:
\begin{eqnarray*}
 \zeta^\Bo &=& 2(\og^\Bo){}^{AB}{(\overline{\Gamma}^\Bo)}{}^{r_\Bobo}_{AB} +  \tau^\Bo (\og^\Bo)^{r_\Bobo r_\Bobo}
\\
 & = &
   2(\og^\Bo)^{AB}\Big( \frac{\partial r_\Bo }{\partial  x^k} \frac{\partial  x^i}{\partial x_\Bo ^A}\frac{\partial  x^j}{\partial x_\Bo ^B}{\overline{\Gamma}}{}^k_{ij} +  \frac{\partial r_\Bo }{\partial r} \frac{\partial^2 r}{\partial x_\Bo ^{A}\partial x_\Bo ^B}\Big)
\\
 && +   \tau\frac{\partial r}{\partial r_\Bo } \frac{\partial r_\Bo }{\partial x^{i}}\frac{\partial r_\Bo }{\partial x^{j}}{\og}{}^{ij}
\\
&=&
  \frac{\partial  r_\Bo }{\partial  r}\zeta
 +  2  \frac{\partial r_\Bo }{\partial r}\Delta_{{\check g}}r
 + O(r_{\Bo}^{-3})
\;,
\end{eqnarray*}
where $\Delta_{\check g}$ is the Laplace operator of the two-dimensional metric $\check g_{AB}dx^A dx^B$.
From this one easily obtains
\bel{30XII13.1}
  \zeta^\Bo_2  = \zeta_2 +\tau_2 +   \Delta_s \tau_2
\;.
\end{equation}
Inserting \eq{30XII13.1} into \eq{29XII13.22} one obtains \eq{29XII13.11}, which completes the proof.

In fact, the calculations just made show that the integral \eq{29XII13.11} is invariant under changes of the coordinate $r$ of the form $r\mapsto r + r_0(x^A) + O(r^{-1})$.

Let $S$ be a section of $\scrip$ arising from a smooth light-cone as above, and let $S'$ be any section of $\scrip$ contained entirely in the past of $S$.
The Trautman-Bondi mass loss formula shows that
$\mTB(S')$ will be larger than or equal to $\mTB(S)$~\cite[Section~8.1]{JAPP}. So, our formula \eq{29XII13.16+} establishes positivity of $\mTB$ for all such sections $S'$.

In many cases the past limit of $\mTB$ is the ADM mass, and one expects this to be true quite generally for asymptotically Minkowskian space-times. Whenever this and  \eq{29XII13.16+}
hold, we obtain an elementary proof of non-negativity of the ADM mass.

It is tempting to use a density argument to remove the hypothesis of non-existence of conjugate point precisely at $\scrip$ along some generators; such generators will be referred to as \emph{asymptotically singular}.  For this one could first rewrite \eq{29XII13.16+} as
\bean
&& \mTB
  = \frac 1 {  16 \pi}
 \int_0^\infty  \int_{S^2} \bigg(  \half   |\xi|^2  + S
\\
 && \phantom{x}\,
   +
(  |\sigma|^2
    +
  8 \pi T_{rr}|_{\mcN}
   )e^{   \int_r^\infty \frac {\tilde r\tau-2} {2 \tilde r} \,d\tilde r  } \bigg)
   r^2 e^{ -  \int_r^\infty \frac {\tilde r\tau-2} {  \tilde r} \,d\tilde r  }
   d\mu_{s} dr
   \!\!
 \;,
 \phantom{xxx}
\eeal{29XII13.16++}
where we have used \eq{volume_element}.
One might then consider an increasing sequence of tensors $\sigma_i$ which converge to $\sigma$ as $i\to\infty$ so that $|\sigma_i|^2$ converges to $|\sigma|^2$ from below, and such that for each $i$ the associated solution $\tau_i$ of the Raychaudhuri equation leads to a metric satisfying \eq{Riemannian}. 
It is easy to see that $\tau_i$ is then decrasing to zero along the asymptotically singular generators as $i$ tends to infinity, monotonically in $i$, which leads to an infinite integral $ -  \int_r^\infty \frac {\tilde r\tau-2} {  \tilde r} \,d\tilde r$  on those generators. It is however far from clear if and when this divergence leads to a finite volume integral after integrating over the generators, and we have not been able to conclude along those lines.

Suppose, finally, that instead of a complete light-cone we have a smooth characteristic hypersurface $\mcN$ with an interior boundary $S_0$ diffeomorphic to $S^2$, and intersecting $\scrip$ transversally in a smooth cross-section $S$ as before. Let $r$ be an affine parameter on the generators chosen so that $S_0=\{r=r_0\}$ for some $r_0 > 0$   and such that \eq{29XII13.3}-\eq{29XII13.4} hold. The calculations above give the following formula for $\mTB$:
\begin{equation}
 \mTB   =   \mbox{\rm r.h.s. of \eq{29XII13.16++}} + \frac 1 {16\pi} \Big(
  8 \pi r_0
 +
   \int_{r=r_0}\big( \zeta  + (  2 {r^{-1}}- \tau)
   {  e^{  \int_{r_0}^\infty \frac {r \tau - 2}{2r} dr }\big) d\mu_{\check g}}
   \Big)
  \;,
\label{2I14.1}
\end{equation}
with the range
$r\in [0,\infty)$ replaced by  $r\in[r_0,\infty)$ in the first integral symbol appearing in \eq{29XII13.16++}. (For outgoing null hypersurfaces issued from any sphere of symmetry in the domain of outer-communications in Schwarzschild, the term multiplying the exponential in \eq{2I14.1}  and the right-hand side of \eq{29XII13.16++} vanish, while the remaining terms add to the usual mass parameter $m$.)

\Eq{2I14.1} leads to the following interesting inequality for \emph{space-times containing white hole regions}.
  We will say that a surface $S$ is \emph{smoothly visible} from $\scrip$ if the null-hypersurface generated by the family of outgoing null geodesics is smooth in the conformally rescaled space-time. Assume, then,  that $S$ is smoothly visible and  \emph{weakly past outer trapped}:
\bel{29III14.1}
 \zeta|_S \ge 0
 \;.
\ee
Smooth visibility implies that $\tau>0$ everywhere, in particular on $S$. By a translation of the affine parameter $r$ of the generators of $\mcN$  we can achieve
\bel{29III14.2}
 r_0 = \frac{ 2}{\sup_S \tau}
 \;.
\ee
All terms in \eq{2I14.1} are non-negative now,
leading to the interesting inequality
\bel{29III14.2}
  \mTB \ge  \frac{ 1}{\sup_S \tau}
 \;.
\ee

Further, equality implies that $S$ and $T_{rr}$ vanish along  $\mcN$, and that we have
\bel{29III14.3}
 \xi=\sigma=0 \ \mbox{along}\  \mcN\;, \quad   \zeta|_S=0
 \;,
 \quad
 \tau|_S\equiv\frac 2 {  r_0} = \frac 1 {\mTB}
 \;.
\ee
Integrating the Raychaudhuri equation gives
\bel{29III14.4}
 \tau=\frac2 {  r }\;,
 \quad
 r\ge r_0
 \;.
\ee
The equations $ \tau  \ol g_{AB}=2 \chi_{AB}=   \partial_r \ol g_{AB}$ together with the asymptotic behaviour of the metric  imply
\bel{29III14.5}
 g_{AB}= r^2 s_{AB}
 \;,
 \quad
 r\ge r_0
 \;.
\ee
It follows that the outgoing null hypersurface issued from $S$ can be isometrically embedded in the Schwarzschild space-time with $m=\mTB$ as a null hypersurface emanating from a spherically symmetric cross-section of the past event horizon.

  Time-reversal and our result provide, of course, an inequality between the mass of the  past directed null hypersurface issuing from a weakly future outer trapped surface $S$ and $\sup_S \zeta$ in black hole space-times.

We finish this paper by noting that formulae such as 
\eq{2I14.1}, together with monotonicity of mass,  might be useful in a stability analysis of black hole solutions, by chosing the boundary  to lie on the initial data surface, for then one obtains an a priori $L^2$-weighted bound on $|\sigma|^2$ and $|\xi|^2$ on all corresponding null hypersurfaces.

\bibliographystyle{amsplain}

\bibliography{../references/hip_bib,%
../references/reffile,%
../references/newbiblio,%
../references/newbiblio2,%
../references/chrusciel,%
../references/bibl,%
../references/howard,%
../references/bartnik,%
../references/myGR,%
../references/newbib,%
../references/Energy,%
../references/dp-BAMS,%
../references/prop2,%
../references/besse2,%
../references/netbiblio,%
../references/PDE}

\end{document}